# Categorization of Mechanics Problems by Students in Large Introductory Physics Courses: A Comparison with the Chi, Feltovich, and Glaser Study

Andrew Mason[1], and Chandralekha Singh[2]

1. Department of Physics and Astronomy, University of Central Arkansas, Conway, AR 72035
2. Department of Physics and Astronomy, University of Pittsburgh, Pittsburgh, PA 15260

**Abstract:** With inspiration from the classic study by Chi, Feltovich, and Glaser [1], we asked introductory physics students in three introductory physics classes to categorize mechanics problems based upon similarity of solutions. To evaluate the effect of problem context on students' ability to categorize, two sets of problems were developed for categorization. Some problems in one of the problem sets that students were asked to categorize included those available from the prior study by Chi et al. Our findings, which contrast from those of Chi et al., suggest that there is a much wider distribution of expertise among introductory students.

## INTRODUCTION

Categorizing or grouping together problems based upon similarity of solution is often claimed to be a hallmark of expertise [1-5]. An expert in physics may categorize many problems involving conservation of energy in one category and those involving conservation of linear momentum in another category, even if some of the problems involving the different conservation laws may have similar contexts and other problems involving conservation of energy only may have very different contexts.

In the classic study conducted by Chi, Feltovich and Glaser [1] (here called the Chi study), eight introductory physics students in calculus-based courses were asked to categorize introductory mechanics problems based upon similarity of solution. Unlike experts who categorize problems based on the physical principles involved in solving them, introductory students were sensitive to the contexts or "surface features" of the problems and categorized problems involving inclined planes in one category and pulleys in a separate category [1]. On the other hand, physics graduate students (experts) were able to identify physics principles applicable in a situation and categorize the problems based upon those principles.

With inspiration from the Chi study [1], we compare the categorization of physics problems by students in large calculus-based and algebra-based introductory courses. Within the theoretical framework that expert and novice categorizations differ, we investigate a potentially wider spectrum in students' expertise in large introductory classes than was possible to capture by analyzing data from only 8 introductory student volunteers in the Chi [1] study.

Although full comparison is impossible without access to all of the original problems in the Chi study, we compare the distribution of students' expertise in categorizing problems in introductory physics classes with the eight introductory student volunteers in the Chi study. We use two versions of the problem set, one of which involved the Chi problems available. We investigate whether there is a qualitative difference between the 8 Chi students and our student populations. Due to the small sample size in the Chi study, we refrain from performing statistical analysis of the Chi data on the grounds that the standard error will be too large to determine anything meaningful. To evaluate the effects of problem topic and of context within a mechanics topic on students' ability to categorize, two sets of problems were developed for categorization. Problems in version II included all seven problems available from the Chi study.

## METHODOLOGY

All students who performed the categorization task were provided the following instruction at the beginning of the problem set: *Your task is to group the 25 problems below based upon similarity of solution into various groups on the sheet of paper provided. Problems that you consider to be similar should be placed in the same group. You can create as many groups as you wish. The grouping of problems should NOT be in terms of ``easy problems'', ``medium difficulty problems'' and ``difficult problems'' but rather it should be based*

*upon the features and characteristics of the problems that make them similar. A problem can be placed in more than one group created by you. Please provide a brief explanation for why you placed a set of questions in a particular group. You need NOT solve any problems.*

The sheet on which participants were asked to perform the categorization of problems had three columns. The first column asked them to use their own category name for each of their categories, the second column asked them for a description of each category that explains why problems within that category may be grouped together, and the third column asked them to list the problem numbers for the questions that should be placed in a category. Apart from these directions, neither students nor faculty were given any other hints about which category names they should choose.

As a preliminary check to make sure the problems were clear, we conducted individual interviews with a few introductory students and physics professors in which they were asked to categorize the problems using think-aloud protocol, and we found that all of them interpreted the instructions as intended. One difference between this study and the Chi study is that, since few students were involved in the Chi study, students were given each of the problems on index cards that could be sorted and placed in groups. In our study, which involved hundreds of students, the categorization task was necessarily a paper-and-pencil task requiring students to write down their reasoning as well as their categories. Based upon the nature of the task, we do not anticipate that the performance of an individual with a certain level of expertise in mechanics as manifested by categorization of problems will be significantly affected by either of these implementation strategies.

Furthermore, in the Chi study, a record of how much time each student took to perform categorizations was maintained. In an in-class study with a large number of students, it was not practical to keep track of time. Instead, all students had 50 minutes to perform the categorization.

Each version of the problem set contained 25 mechanics problems, 15 of which were included in both sets. The remaining 10 were unique for each problem set. The context of the 25 problems varied. Only 7 problems (called Chi problems for convenience) from the Chi study were known to us because they were the only ones mentioned in the Chi study and thus identifiable by the problem numbers from the third edition of the introductory physics textbook by Halliday and Resnick (1974 edition). Personal communication with the lead author of Ref. [1] suggested that the problems in the original study not mentioned in their paper had been discarded and were not available. In Version I, which did not include any of the Chi problems, all of the 25 problems were developed by us. The problems were on sub-topics similar to those chosen in the Chi study. The topics included one- and two-dimensional kinematics, dynamics, work-energy theorem, and impulse-momentum theorem and were distributed among these different topics as evenly as possible. Version II, which included the 7 Chi problems, also had 3 non-Chi questions on rotational kinematics and dynamics and angular momentum. The purpose of including additional (non-Chi) problems on rotational motion was an attempt to match these questions to the related Chi questions by deep structure, and thus eliminate the possibility that the Chi questions would stand out by being the only questions dealing with rotational kinematics and dynamics. Chi problems were included in Version II in order to evaluate how students performed on those problems compared to the non-Chi problems. Version I had 10 problems that were different from the 7 Chi problems and the 3 problems on rotational motion.

The 7 Chi questions used in this study involved non-equilibrium applications of Newton's laws, rotational motion or the use of two physics principles. While some of our problems also covered the same topics and had similar features, it is impossible to predict the exact match with other topics covered by other Chi problems (not available) even though they were also from mechanics. In choosing our own problems, we tried to cover the topics from the chapters in introductory mechanics and we also included problems with various levels of difficulty. For example, two-part problems or non-equilibrium problems are more challenging than one-part problems or equilibrium problems.

Two algebra-based classes with 109 and 114 students and one calculus-based class with 180 students performed the categorization in their recitation classes. All relevant concepts in the categorization problem sets were taught prior to administration of the task. All students were told that they should try their best but they were given the same bonus points for doing the categorization regardless of how expert-like their performances were. One algebra-based class (with 109 students) was given Version II which included Chi problems.

| Chi's Categories | % of 1981 novices (8 total) | % of 1981 experts (8 total) | % of algebra-based students version II (109 total) | | % of algebra-based students version I (114 total) | % of calculus-based students (180 total) |
|---|---|---|---|---|---|---|
| | | | All questions (25) | Chi questions (7) | | |
| Novice Categories from the Chi Study | | | | | | |
| Angular motion (including circular) | 87.5 | - | 72 | 59 | 57 | 42 |
| Inclined planes | 50 | - | 24 | 19 | 19 | 18 |
| Velocity and acceleration | 25 | - | 31 | 26 | 51 | 10.5 |
| Friction | 25 | - | 55 | 51 | 52 | 27 |
| Kinetic energy | 50 | - | 16 | 15 | 15 | 6 |
| Cannot classify/omitted | 50 | - | 44 | 18 | 34 | 39 |
| Vertical motion | 25 | - | 3 | 3 | 3 | 1 |
| Pulleys | 37.5 | - | 16 | 16 | 6 | 2 |
| Free fall | 25 | - | 6 | 1 | 4 | 6 |
| Expert Categories from the Chi Study | | | | | | |
| Newton's 2nd Law (also Newton's Laws) | - | 75 | 22 | 18 | 19 | 38 |
| Energy principles (conservation of energy, work-energy theorem, energy considerations) | - | 75 | 42 | 31 | 35 | 73 |
| Angular motion (not including circular) | - | 75 | 43 | 31 | 39 | 15 |
| Circular motion | - | 62.5 | 29 | 28 | 18 | 27 |
| Statics | - | 50 | 0 | 0 | 0 | 0 |
| Conservation of Angular Momentum | - | 25 | 7 | 1 | 1 | 1 |
| Linear kinematics/motion (not including projectile motion) | - | 25 | 51 | 44 | 42 | 63 |
| Vectors | - | 25 | 1 | 1 | 16 | 2 |
| Categories made by both Novices and Experts from the Chi Study | | | | | | |
| Momentum principles (conservation of momentum, momentum considerations) | 25 | 75 | 39 | 11 | 33 | 64 |
| Work | 50 | 25 | 4 | 4 | 41 | 47 |
| Center of mass | 62.5 | 62.5 | 2 | 0 | 1 | 0 |
| Springs | 75 | 25 | 23 | 23 | 52 | 30 |

**TABLE 1**. Performance in our study vs. performance in the Chi study. The novice and expert categories are those made by introductory physics students and graduate students, respectively, in the Chi study. Introductory physics students in the calculus-based courses (last column) were much more likely than those in the algebra-based courses to place problems in expert-like categories such as Newton's Second Law, Energy principles, Linear Kinematics, Momentum principles and Work. Categories shaded in gray are those for which the questions from the Chi study were not available. Our questions seldom belonged to those categories.

## RESULTS

Classification of categories consisted of placing each category created by each student into a matrix with problem numbers along the columns and categories along the rows. A "1" (or "0") was placed in a box if the problem appeared (or did not appear) in the given category. For those who categorized Version II, an average of 7.02 categories per student was created. We recorded 82 proto-categories, which were later reinterpreted into 59 categories.

Table 1 shows the list of categories that "experts" and "novices" created in the Chi study. The Table includes the percentages of both novices and experts in the Chi study, and students in the calculus-based and two algebra-based courses in our study. For Version II, which included 7 Chi problems, two separate columns in Table 1 show categorization for only those seven questions and all questions.

Table 1 shows that the percentage of introductory students in our study who selected "inclined planes" or "pulleys" as categories (based mainly upon the surface features) is significantly less than in the Chi study. One possible reason could be that there were fewer questions involving ramps and pulleys than in the Chi study. However, Table 1 shows that "ramp" was also a much less popular category for introductory students in our study for Version II, in which 19% chose this category for Chi problems and 24% for non-Chi problems (compared to the Chi study, in which 50% of the students placed at least one problem in this category). Similarly, Table 1 shows that kinetic energy was a novice category selected by 50% of the Chi students but in our study using both versions it was never more than 16%.

What is more surprising however is that none of the 8 introductory students in the Chi study (see Table 1) chose Chi's expert categories, e.g., Newton's second law, energy principles, circular motion or linear kinematics as categories at all. On the other hand, Table 1 shows that in our study with Version II, 18% selected Newton's second law for the 7 Chi-problems (22% for all), 31% selected energy principles for the 7 Chi problems (42% for all), 28% selected circular motion for the 7 Chi problems (29% for all) and 44% selected linear kinematics for the 7 Chi problems (51% for all). The fact that there were absolutely no introductory students choosing these categories in the Chi study (see Table 1) but the percentage of students selecting these categories is quite large, *even for the Chi problems used in our study*, is hard to reconcile even considering the small number of student volunteers in the Chi study. One factor contributing to this large difference may be that the student volunteers in the Chi study may not have currently been taking the course (and may have forgotten the material), while the students in this study were concurrently enrolled in an introductory physics course. Further, we note that Version II was only given to algebra-based introductory students who are generally likely to be worse at performing expert-like categorizations than the calculus-based introductory students. Therefore, the large discrepancies between the expert-like categorizations in our study and the Chi study are likely to be even larger if we had given the 7 Chi-problems to the calculus-based group.

## SUMMARY

It is striking that while none of the introductory students in the Chi study selected "expert" categories such as "Newton's second law", or "linear kinematics", a significant number of introductory students chose these categories in our study. Even if we restrict our study to the 7 problems common with Chi, Table 1 shows that the number of introductory students who selected such "expert" categories is significantly larger than zero in the Chi study. While it is not possible to compare our other data directly with that in the Chi study (most of their questions are not available), the percentage of introductory physics students who chose "surface-feature" categories such as "ramp" and "pulley" was significantly lower than the percentages reported in the Chi study. The distribution of students' expertise in an in-class study like ours at a typical state university is likely to reflect the distribution in a typical classroom (including high achieving and low achieving students). On the other hand, the distribution in an out of class study is more unpredictable and depends on the volunteer pool including issues such as how long ago they took the physics course. Our finding suggests that since expertise plays a role in categorization, it is not appropriate to call all introductory students novices as in the Chi study.